\documentclass[fleqn,10pt]{wlscirep}
\usepackage[utf8]{inputenc}
\usepackage[T1]{fontenc}
\usepackage{subfig}

\title{Non-Compulsory Measures Sufficiently Reduced Human Mobility in Tokyo during the COVID-19 Epidemic}

\date{April 20th, 2020}

\author[1]{Takahiro Yabe}
\author[2]{Kota Tsubouchi}
\author[3,4,5]{Naoya Fujiwara}
\author[6]{Takayuki Wada}
\author[4]{Yoshihide Sekimoto}
\author[1,*]{Satish V. Ukkusuri}
\affil[1]{Lyles School of Civil Engineering, Purdue University, West Lafayette, Indiana, USA}
\affil[2]{Yahoo Japan Corporation, Tokyo, Japan}
\affil[3]{Graduate School of Information Sciences, Tohoku University, Sendai, Japan}
\affil[4]{Institute of Industrial Science, the University of Tokyo, Tokyo, Japan}
\affil[5]{Center for Spatial Information Science, the University of Tokyo, Kashiwa, Japan}
\affil[6]{Graduate School of Human Life Science, Osaka City University, Osaka, Japan}

\affil[*]{sukkusur@purdue.edu}


\begin{abstract}
While large scale mobility data has become a popular tool to monitor the mobility patterns during the COVID-19 pandemic, the impacts of non-compulsory measures in Tokyo, Japan on human mobility patterns has been under-studied. 
Here, we analyze the temporal changes in human mobility behavior, social contact rates, and their correlations with the transmissibility of COVID-19, using mobility data collected from more than 200K anonymized mobile phone users in Tokyo.  
The analysis concludes that by April 15th (1 week into state of emergency), human mobility behavior decreased by around 50\%, resulting in a 70\% reduction of social contacts in Tokyo, showing the effectiveness of non-compulsory measures. 
Furthermore, the reduction in data-driven human mobility metrics showed correlation with the decrease in estimated effective reproduction number of COVID-19 in Tokyo. 
Such empirical insights could inform policy makers on deciding sufficient levels of mobility reduction to contain the disease. 
\end{abstract}

\begin{document}

\flushbottom
\maketitle

\thispagestyle{empty}



\section*{Introduction}
The COVID-19 pandemic has posed unprecedented challenges for cities around the globe \cite{anderson2020will}. 
Countries have tackled this challenge with a variety of non-pharmaceutical interventions (NPIs), ranging from complete regional lockdowns, closures of non-essential businesses, to testing and tracing \cite{flaxman2020report}. 
In order to monitor, analyze and evaluate the impacts of such interventions, large scale mobile phone data has been identified as an effective data source \cite{oliver2020mobile}. 
Previous studies on human mobility analysis have shown that such large-scale mobility data collected from mobile devices (e.g. GPS, call detail records) may be used to assist the modeling of epidemic spread \cite{bengtsson2015using,finger2016mobile,tizzoni2014use,wesolowski2012quantifying}.
During the current COVID-19 crisis, researchers from academia, industry, and government agencies have started to utilize large-scale mobility datasets to estimate the effectiveness of control measures in various countries including China, Germany, France, Italy, Spain, Sweden, United Kingdom and the United States \cite{klein2020assessing,kraemer2020effect,lai2020effect,pepe2020covid,bonato2020mobile,wellenius2020impacts,gao2020mapping,dahlberg2020effects,santana2020analysis,cintia2020relationship}.

Several studies have been conducted to analyze the mobility patterns in Japan during the state of emergency due to COVID-19 \cite{mizuno}.
Moreover, private companies such as  Agoop \cite{agoop} and NTT Docomo \cite{nttdocomo} have visualized changes in population distributions during the spread of COVID-19 using mobile phone data. 
However, we lack studies that attempt to analyze the mobility behavioral changes in Japan and to further understand its correlation with the spread of COVID-19. 
Japan, in particular, has experienced a significantly low number of cases and deaths due to COVID-19 in comparison to other countries in Europe and America, despite the social and physical proximity to China and intervention policies that are not as aggressive as some of the other countries \cite{dong2020interactive,idogawa}. 
NPIs implemented by the Japanese government include non-mandatory closures and remote-working of non-essential business employees (February 26th), closures of public elementary, junior high and high schools (March 2nd), and incremental inbound entry restrictions, starting with personnel who have visited Hubei Province, China (February 3rd), until restricting inbound visitors from 73 countries (April 3rd). 
In the case of Japan, the current justice system does not allow the government to lay out a mandatory lockdown. 
Such differences in policies and COVID-19 spread dynamics make Japan an interesting case study for international comparative analysis. 
The objective of this paper is to provide empirical analysis on the effects of such non-compulsory NPIs on changes in i) individual human mobility behavior and ii) amount of social contacts, and to iii) analyze the correlations between such mobility changes and transmissibility of COVID-19.  
In this study, large-scale mobility data collected from more than 200K mobile phones across a four month period in Tokyo are used to answer the aforementioned research questions. 

\begin{figure*}[t]
\centering
\subfloat{\includegraphics[width=\linewidth]{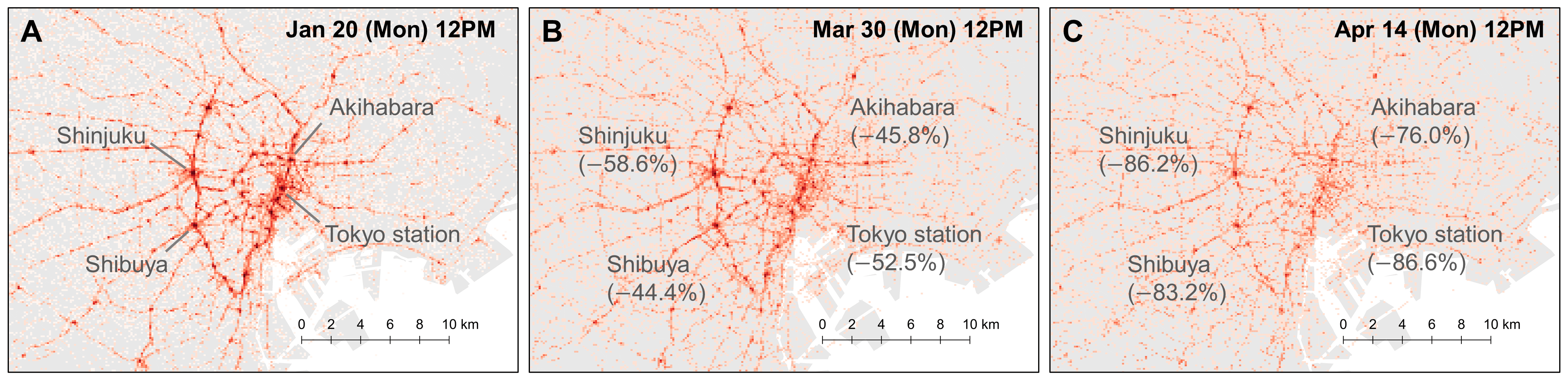}} \\
\subfloat{\includegraphics[width=\linewidth]{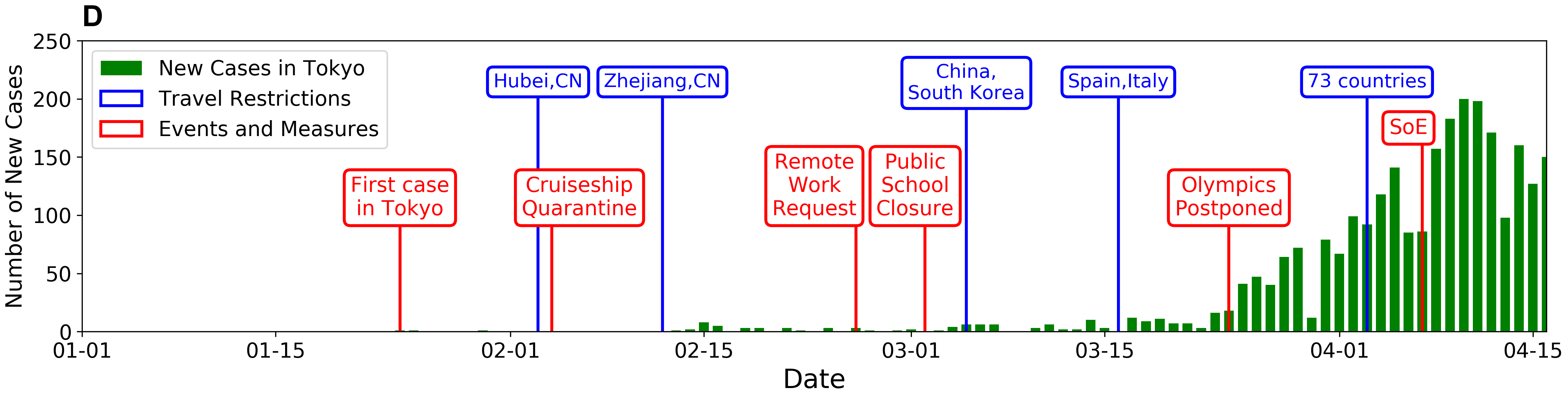}}
\caption{\textbf{Macroscopic mobility dynamics.} (A)-(C) show the population distributions on 3 different dates at same times (12PM), each on the same day of week (Mondays). Significant decrease in the population density at stations and cities along the Yamanote-line (ring railway) can be observed. (D) shows the timeline of the COVID-19 response and occurrence of new cases in Tokyo. Green bars show the number of cases for each day, blue annotations show inbound travel restrictions, and red annotations list related events and the policy interventions by the Japanese Government.}
\label{fig:timeline}
\end{figure*}

\section*{Results}
\subsection*{Macroscopic Mobility Dynamics and Timeline}
Panels A-C in Figure \ref{fig:timeline} show the daytime population distributions on 3 different dates at same times (12PM), each on the same day of week (Mondays) in Tokyo, Japan.
Mobile phone data were interpolated to produce movement trajectories, and the locations of the users at 12PM on each day were spatially aggregated into 100 meter grid cells. 
The population density was corrected by the number of observed users on each day, as the number of active users had daily fluctuations. 
As shown in previous studies \cite{kashiyama2017open}, we observe high population density along the railway lines and hub stations during the daytime. 
Comparing the three panels, we observe significant decrease in population density at stations and cities along the Yamanote-line (ring railway), such as Shibuya, Shinjuku, Akihabara, and Tokyo Stations. 
By April 14th, which was 1 week after the declaration of state of emergency, these hub stations had a 76\% to 87\% decrease in foot traffic compared to the baseline, which was computed by taking the average population density of weekdays at 12PM before the crisis (January 2020). 

Figure \ref{fig:timeline}D shows the timeline of the COVID-19 response and occurrence of cases in Tokyo Metropolis.
Green bars show the number of cases for each day, blue annotations show inbound travel restrictions, and red annotations list the policy interventions by the Japanese Government.
The number of cases for each day were obtained from the Tokyo Metropolitan Government website \cite{tokyometro}.
Major non-pharmaceutical measures taken by the government include issuing non compulsory requests for remote working to private companies on February 26th, and closing public schools (elementary, junior high, and high schools) from March 2nd until the end of the semester, which is usually the end of March \cite{mhlw}. 
The 2020 Olympic Games were postponed for a year on March 24th, and the State of Emergency (SoE) was declared on April 7th. 
Entry restrictions for inbound travellers were gradually reinforced, with the first restriction on February 3rd from Hubei Province, China. 
The restrictions were expanded to all foreign nationals who visited China or South Korea during the past 14 days on March 5th, and to 73 countries including the United States on April 3rd.

\subsection*{Changes in Human Mobility Behavior}
\begin{figure*}[t]
\centering
\includegraphics[width=\linewidth]{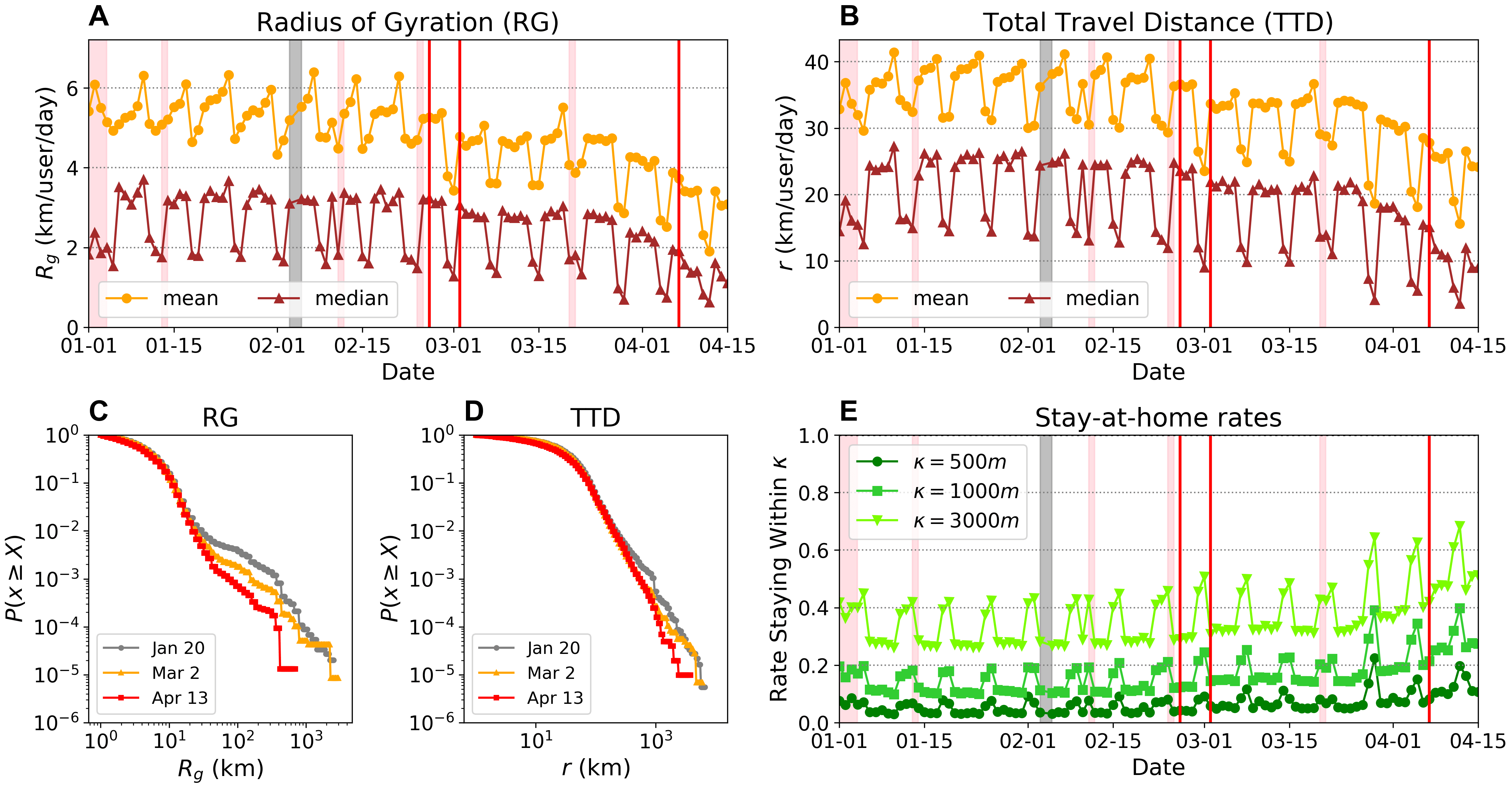}
\caption{\textbf{Changes in individual mobility patterns.} Temporal transition of mean and median radius of gyration (A), total travel distance per day (B) and their complementary cumulative distribution functions (C), (D), respectively. Panel (E) shows the rate of staying $\kappa$ meters within estimated home locations.}
\label{fig:indiv}
\end{figure*}

Here, we evaluate how individual users' mobility behavior has changed before and after the state of emergency in Tokyo, Japan.
Figure \ref{fig:indiv} shows the transitions of various metrics of individual mobility patterns.
Panels A and B in Figure \ref{fig:indiv} show the mean and median values of the radius of gyration (RG) and total travel distance (TTD), which characterize the magnitude and extent of individual mobility behavior \cite{gonzalez2008understanding} (see \textbf{Methods}). 
We observe two steps of decrease in both metrics, the first starting around the end of February and stabilizing until late March, and the second which is a more rapid decrease until April 15th. 
Both the mean and median values of RG and TTD have been reduced by 50\% of the typical values by the end of the observation period on April 15th. 
Panels C and D in Figure \ref{fig:indiv} show the complementary cumulative probability distribution plots for both mobility metrics across three different dates, January 20th (before the pandemic), March 30th (before the state of emergency), and April 14th (after the state of emergency), which are all Mondays. 
The RG values observe a dip at around $20$km, which corresponds to the diameter of the Tokyo Metropolitan area (scale shown in Figure \ref{fig:timeline}A-C). 
This implies that while the majority of the individuals move within the metropolitan area, some individuals travel to different regions outside Tokyo.
In both mobility metrics, we observe significant tempering of the tail probability as time passes.
This indicates a decrease in relatively long distance travelers ($>20$km for RG and $>100$km in TTD), reflecting the decrease and closures of airlines and inter-urban railways that travel outside the metropolitan Tokyo area. 
Figure \ref{fig:timeline} E shows the rates of users staying $\kappa$ meters within their estimated home locations. 
This figure suggests that the rate of people staying close to their estimated home locations increased since mid-March and doubled by April 15th regardless of the threshold value $\kappa$, reflecting the increase in people working from home and spending time within their neighborhoods.

\subsection*{Social Contact Analysis}
\begin{figure*}[t]
\centering
\includegraphics[width=\linewidth]{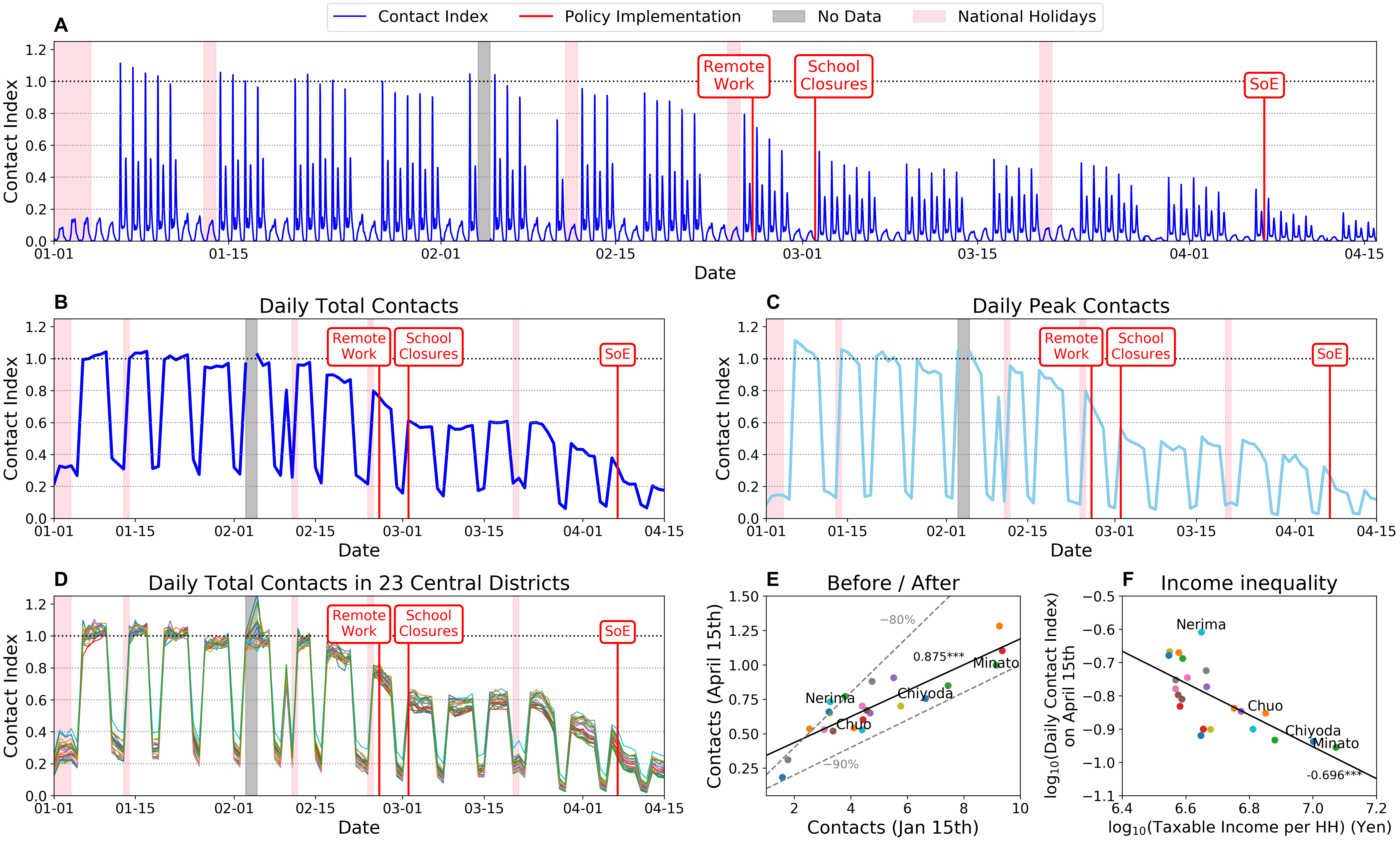}
\caption{\textbf{Social contact analysis.} (A) shows the amount of contacts an individual encounters outside home for each time period. (B) and (C) show the individuals' mean total daily contacts and the mean daily peak contacts, respectively. (D) Daily total contacts in the 23 central districts (wards) in Tokyo Metropolitan Area. (E) Correlation between non-normalized social contacts before and after the COVID-19 spread. (F) Correlation between average taxable income per household and daily contact index on April 15th.}
\label{fig:contact}
\end{figure*}

Figure \ref{fig:contact} shows the results of the social contact analysis in Tokyo over the observation period (see \textbf{Methods}). 
The panel figures show the temporal transitions in the relative average social contact index.
The contact index is calculated by tracking the number of individual users that each user encounters in a given temporal window, using the interpolated trajectory data.
We have used 100 meters as the spatial threshold parameter for detecting encounters for this analysis. 
The social contact indexes were normalized by the mean peak contact index values observed prior to the pandemic (January 2020). 
Thus, in Figure \ref{fig:contact}, contact index value of 1 indicates the average amount of observed contacts during peak hours on a typical weekday prior to the pandemic. 
The temporal periodicity in Figure \ref{fig:contact} A shows the daily and weekly patterns, including the highest point showing the morning rush hour peak, the following peak showing the returning home rush hour, and significantly small amount of contacts (around 0.2) on weekends and holidays, including the New Year breaks. 
Panels B and C in Figure \ref{fig:contact} show the total daily contact index and the daily peak contact index values, respectively. 
By comparing the average social contacts across the time horizon, we can observe in Figure \ref{fig:contact} A that the average amount of social contact for each user starts to decrease as COVID-19 became a threat globally in mid February down to 90\% of the typical value. 
The value decreases more rapidly towards the end of February with the government request for remote working, down to 65\% of the typical values. 
The contact index stays at 65\% of typical values until the end of March, and as the number of cases in Tokyo starts to increase (Figure \ref{fig:timeline}), the contact index further starts to decrease and reaches 30\% of typical values by April 15th, almost reaching the government's goal of reducing social contacts by 80\% of the typical values \cite{eighty}.
However, regarding the daily peak contacts (morning rush hour peaks on weekdays and lunchtime peaks on weekends and holidays), Panel C shows that peak time social contacts have decreased successfully to around 20\% of the typical values by April 15th. 

Moreover, the inequality in mobility changes across different spatial regions within Metropolitan Tokyo were tested, since it has been suggested that inequality in society could aggravate the spread of COVID-19 \cite{ahmed2020inequality}.
Panel D of Figure \ref{fig:contact} shows the transition of daily total contacts for the 23 wards in central Tokyo. 
Although the social contact index of the regions follow similar general patterns, some heterogeneity in the contact decrease can be observed. 
Panel E shows strong correlation (Pearson's correlation coefficient $\rho=0.875^{***}$) between the non-normalized amount of contacts on January 15th (before the pandemic) and April 15th, which is after the pandemic. 
Most of the 23 regions were successful in reducing the social contact index by 80\% (shown by gray dashed line), except for wards including Nerima Ward.
Panel F further shows the log-log plot between the relative contact index on April 15th and average taxable income per household for each ward in Tokyo. 
The average taxable income per household was calculated using data provided by the Official Statistics of Japan through the Portal Site \cite{statsjapan}. 
We observe strong negative correlation ($\rho= -0.696^{***}$) between taxable income per household and contact index, indicating that households in higher income regions were able to reduce the amount of social contacts and risk of COVID-19 transmission more than households in lower income regions. This highlights an inequality across households in terms of the voluntary mobility restrictions. Higher income households are more likely able to reduce their social contacts by reducing mobility, however lower income households may not have this flexibility. This inequality has significant implications in terms of disease spread and will require a holistic policy planning. 

\subsection*{Correlation of Mobility Indexes with Effective Reproduction Number $R(t)$}
\begin{figure*}[t]
\centering
\includegraphics[width=\linewidth]{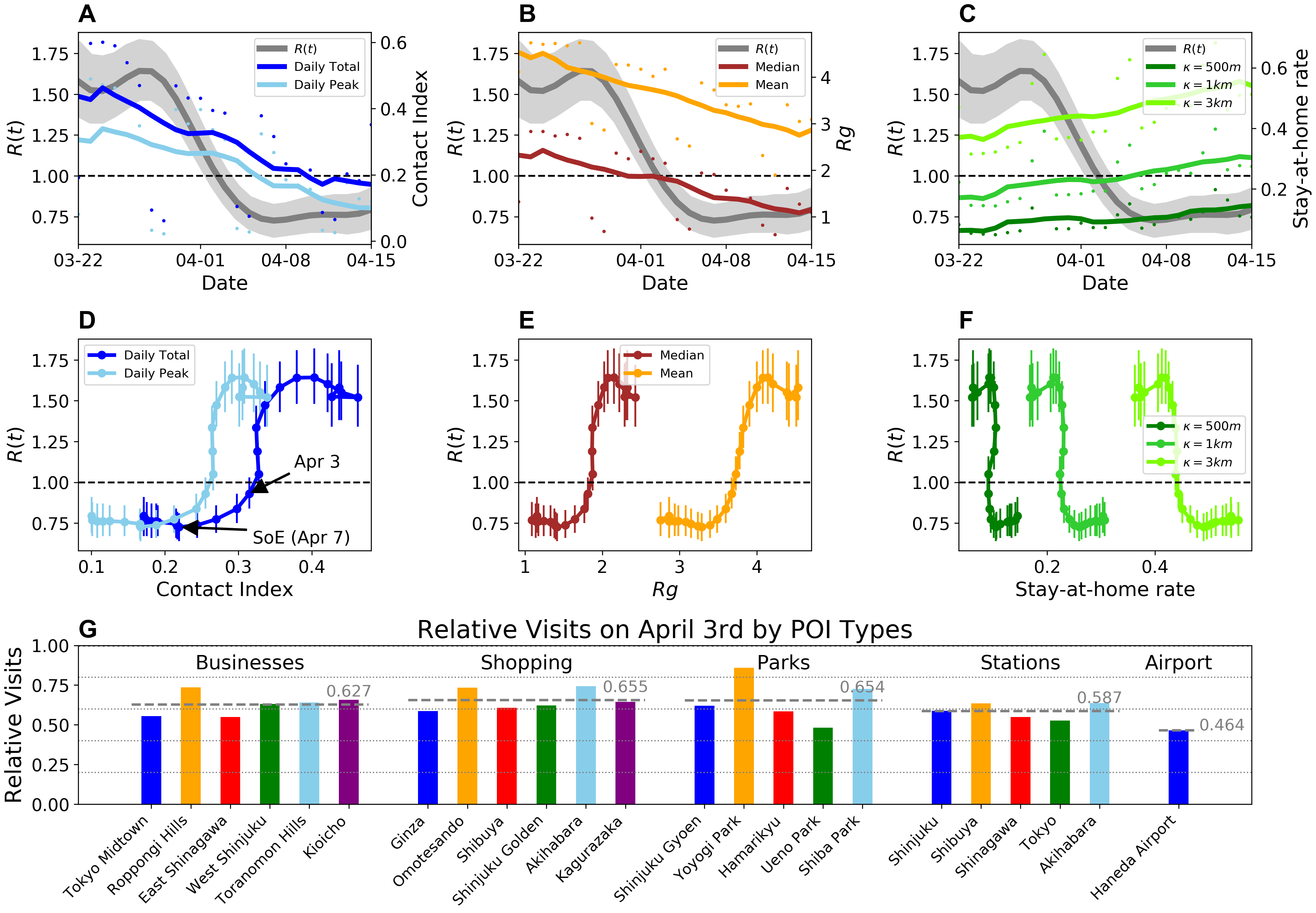}
\caption{\textbf{Correlation between mobility behavior changes and effective reproduction number $R(t)$.} (A-C) shows the human mobility metrics along with the $R(t)$ values in Japan estimated by the health ministry panel on coronavirus infection cluster under the Japanese Government. (D-F) shows the relationships between the mobility metrics and $R(t)$. All panels show that $R(t)$ falls under 1.0 with reduction in mobility and social contacts, and that such reduction may have been excessive in containing the spread of COVID-19. (G) Relative amount of visits to various point of interests on April 3rd, when sufficient reduction in transmissibility was achieved in Tokyo.}
\label{fig:rt}
\end{figure*}

To understand the implications of mobility behavior change and decrease in social contacts on the transmissibility of COVID-19 in Tokyo, we compare the quantified mobility behavior changes with the estimated effective reproduction number $R(t)$ of Tokyo.
The $R(t)$ values were estimated using the weekly averaged number of confirmed COVID-19 cases in Tokyo, reported by the metropolitan government (see \textbf{Methods}).
Figure \ref{fig:rt} shows the weekly moving averaged values of (A) social contact index, (B) radius of gyration, and (C) stay-at-home rates plotted along with the dynamic $R(t)$ values. 
The gray shaded areas show the 95\% confidence interval for the estimated $R(t)$. 
Panels D-F in Figure \ref{fig:rt} show the temporal evolution of the mobility metrics and $R(t)$, where each dot corresponds to daily values. 
The vertical error bars show the 95\% confidence interval of the estimated $R(t)$ values. 
A non-linear relationship between the mobility metrics and transmissibility are observed, where the $R(t)$ values significantly decrease around a threshold value of 0.33 in the contact index. 
We observe that $R(t) < 1$ is achieved with reduction in social contacts down to 0.33 on April 3rd. 
Further reduction in mobility and social contacts are observed after the declaration of the SoE on April 7th, but $R(t)$ had reached a bottom before the declaration.
This suggests that restrictions on mobility certainly allow the reduction of number of transmissions up to a point, but further severe restrictions could have less effects on transmissibility reduction. 
A more targeted and nuanced approach may be needed to determine where and when mobility restrictions should be imposed.

To provide some examples of such nuanced and directly observable metrics of mobility reductions, Panel G shows the relative amount of visits to various types of points of interest (POIs) on April 3rd, which was when sufficient decrease in mobility was observed. 
We can observe that for all listed types of POIs (Table S2) including business districts (e.g. Tokyo Midtown, Roppongi Hills), shopping areas (e.g. Ginza, Omotesando), and major stations (e.g. Shinjuku, Shibuya), visits were reduced by around 35\% to 40\% on average compared to January weekdays (before the pandemic). 
Haneda Airport (major airport in Tokyo area) on the other hand, had larger decrease in visits (55\%) due to travel restrictions which were imposed in early February and March (Figure S7) . 

\section*{Discussion}
Large-scale mobility data collected from mobile phones provide us with an opportunity to monitor and understand the impacts of NPIs during the COVID-19 pandemic with an unprecedented spatio-temporal granularity and scale. 
In this study, we utilized such data to quantify the changes in human mobility behavior and social contacts during the COVID-19 spread in Tokyo, Japan, which provides a unique case study where government policies were limited to non-compulsory measures. 
The analysis concludes that by April 15th (1 week from declaring state of emergency), human mobility behavior have decreased by around 50\% both in terms of radius of gyration and total travel distance per user, and also that social contact index in Tokyo had been reduced by more than 70\% both in terms of daily total contacts and daily peak time contacts following non-compulsory measures including remote working requests to businesses and closures of public schools. 
The analysis results also showed that mobility reduction had already taken place before late March (around 40\% decrease of social contacts), when the number of imported cases of COVID-19 started to increase.
This indicates that even before the actual spread of COVID-19 in Tokyo, and even under only non-compulsory measures, a significant amount of cooperation was provided by the citizens to contain the spread of COVID-19.  
By comparing the mobility analysis results with the effective reproduction number $R(t)$ estimated from the daily confirmed number of cases in Tokyo, we were further able to show correlation between the mobility reduction and decrease in transmissibility of COVID-19. 
During late March, $R(t)$ rose above 1 and the number of cases increased, however, the nation-wide state of emergency triggered further reduction in mobility, resulting in $R(t)$ to drop below 1. 
In fact, it was found that reduction of mobility and social contacts beyond certain thresholds (e.g. social contact index of $0.33$) had little incremental effects on the decrease in $R(t)$.
Looking forward into lifting the state of emergency declaration, these results could inform decision making on how much human mobility and social contact reduction is needed to keep the effective reproduction number of COVID-19 below 1. 
Figure \ref{fig:rt}G provides more directly observable measures of mobility reduction at various points of interest which can be used to interpret the sufficient amount of social contact reduction to contain COVID-19. 

The presented empirical results should be considered in the light of some limitations.
First, the presented analysis was limited to understanding the correlations between mobility and contact reduction and transmissibility.
However, microscopic human behavior, such as sanitizing hands more often or higher rates of wearing face masks, could have affected the reduction in transmissibility. 
Because of such factors other than human mobility, the relationships between mobility reductions and transmissibility could change after lifting the state of emergency. 
Due to the lack of data on $R(t)$ under no interventions, it is not certain whether the $R(t)$ values reach a similar value of that in European countries and the US. 
Additional household surveys and interviews need to be conducted to quantify the effects of such behavioral changes.  
Second, GPS location data collected from mobile phones are prone to spatial errors up to 100 meters. 
This spatial error could have biased the results of the mobility metrics. 
However, sensitivity analysis suggests that the conclusions are not affected by the spatial thresholds selected in this study (Figures S3, S4). 
In addition to the 2\% sample rate, the granularity of the data was not adequate to compute the actual contacts between individuals (i.e. 2 meters distance for all individuals). 
Moreover, it is known that mobile phone data contain biases in age groups. 
It is important to analyze the mobility of the elderly population due to the high mortality rate of COVID-19.
Although similar conclusions were reached when different spatial thresholds were used via sensitivity analysis, additional validation using small scale but precise data could further clarify the approximate social contact measure computed in this study using mobile phone data. 

Future research steps include extending the analysis to other cities in Japan such as Sapporo and Osaka, where different patterns of COVID-19 spread have been observed (Figure S1).
Cross-comparative analysis including multiple regions could yield novel and more generalizable insights on the relationships between mobility pattern changes and COVID-19 transmissibility. 
In addition to the spatial dimension, extending the time period of analysis after April 15th and analyzing the dynamics after the lifting of the interventions could strengthen our understanding between social contact decrease and transmissibility reduction. 
Although this study used aggregated measures to quantify the behavioral changes (e.g. average social contacts), more microscopic analysis of mobility changes could be conducted with additional data.

\section*{Materials and Methods}

\subsection*{Mobile Phone Location Data}
Location data of smartphones were collected by Yahoo Japan Corporation through the disaster alert app in order to send relevant notifications and information to the users. 
The users in this study have accepted to provide their location information. 
The data are anonymized so that individuals cannot be specified, and personal information such as gender, age and occupation are unknown. 
Each GPS record consists of a user's unique ID (random character string), timestamp, longitude, and latitude. 
The data acquisition frequency of GPS locations changes according to the movement speed of the user to minimize the burden on the user's smartphone battery. 
If it is determined that the user is staying in a certain place for a long time, data is acquired at a relatively low frequency, and if it is determined that the user is moving, the data is acquired more frequently. 
The data has a sample rate of approximately 2\% of the population, and past studies suggest that this sample rate is enough to understand the macroscopic urban dynamics, even during an emergency \cite{yabe2020understanding}. 
We selected a panel of users who were active each day in Tokyo metropolitan area before, during and after the COVID-19 pandemic. 
This leads to a sample of about 200k users, with approximately 50 data points per user each day (Figure S2).

\subsection*{Socio-Economic Data}
Socio-economic data of 23 wards in Tokyo Metropolis were obtained from the Portal Site of Official Statistics of Japan \cite{statsjapan}. 
The number of households and total taxable income collected from the residents were available through the Portal Site. 
The taxable income per household values used in the analysis in Figure \ref{fig:contact}F were calculated by dividing the total taxable income by the number of households in each region (ward) (Figure S5, Table S1). 

\subsection*{Effective Reproduction Number $R(t)$}
The effective reproduction number $R(t)$ is commonly used to quantify transmissibility of an infectious disease, which is defined as the average number of secondary cases generated by a single infectious case \cite{held2019handbook}. Effects of NPIs such as social distancing can be measured by the effective reproduction number. In order to estimate $R(t)$ of COVID-19 in Tokyo, we employed time series data of daily confirmed COVID-19 cases in Tokyo reported by Tokyo metropolitan government  \cite{tokyometro}. Since the data of the onset cases are unavailable, we used the confirmed date for the estimation. The original daily data were smoothed by taking the centered 7-day moving average.
Estimation of $R(t)$ is conducted based on the code implemented by Jung et al., which is available on the Github repository \cite{reff}. 
The code estimates $R(t)$ using the following steps. First, the backcalculation is performed   \cite{becker1991method} in two steps. 
Since our data is based on the laboratory confirmation, there is a time lag associated with the date of onset. The time duration between the onset of the infections and the actual reported data is estimated using the right truncated Weibull distribution with the shape factor $1.741$ and scale factor $8.573$. 
Next, the date of the infection is estimated with back calculation using the lognormal distribution for the incubation period.
Parameter values of the mean 5.6 days and the standard deviation 3.9 days for COVID-19 estimated by Linton et al. \cite{linton2020incubation} is adopted. 
Finally, $R(t)$ is estimated based on the renewal equation \cite{diekmann2012mathematical}, which relates the time series of the number of new infected cases and the effective reproduction number.
This is done by maximizing the likelihood function based on the assumption that the infection rates follow a Poisson distribution.
To model the serial interval, the Weibull distribution with shape factor $2.305$ and the scale factor $5.452$ is assumed \cite{nishiura2020serial}.


\subsection*{Individual Mobility Indexes}
Several indexes have been used in previous studies to characterize individual human mobility patterns \cite{gonzalez2008understanding,blondel2015survey}.
In this study, we use three key indexes: Radius of Gyration ($R_g$), total travel distance ($TTD$), and stay-at-home rates parameterized by a spatial threshold $\kappa$ ($SAH_\kappa$).
Given a sequence of GPS observation points $P_i = \{ p_i^1,p_i^2, \cdots, p_i^N \}$ in a single day of an individual user $i$, the radius of gyration is calculated using the following equation.
\begin{equation}
    R_g = \sqrt{\frac{1}{N} \sum_{n=1}^N (p_i^n - \Bar{p_i})^2}
\end{equation}
where $\Bar{p_i}$ denotes the center of mass of the GPS observation points of user $i$. 
The total travel distance of an individual user is the sum of Euclidean distances between all subsequent pairs of GPS observation points on a given day. 
The stay-at-home rate, parameterized by a spatial threshold $\kappa$, is the rate of individual users who had stayed the entire day within a distance $\kappa$ from the estimated home location.

\subsection*{Quantifying Social Contact Index from Trajectory Data}
To overcome data sparsity, spatio-temporal interpolation of the GPS location observations were performed. 
Because the GPS data are collected less frequently when movement is detected, we assume that the individual users are static while there are no observations. 
Using the interpolated individual trajectory data produced from mobile phones, the social contact indexes were computed. 
The social contact indexes shown in Figure \ref{fig:contact} were computed for 30 minute intervals. 
First, for each time interval $[t,t+dt)$, where $dt=30$ minutes, users who were not within 100 meters from their estimated home locations were detected as ``staying out''.
We denote this set of individual users as $N^{\rm out}_t$. 
For user $i$ staying out ($i \in N^{\rm out}_t$), we compute the number of other ``staying out'' users who are within 100 meters from user $i$, and use that count $c_{i,t}$ as a proxy for social contacts. 
The social contact index is calculated as the total social contacts for all users staying out, divided by the total number of users including those staying at their homes. 
Thus, mean social contact value is computed as $C_t = 
\sum_{i\in N^{\rm out}_t} c_{i,t}/N$, where $N$ is the total number of users observed on that day. 
The social contact index is the relative value of mean social contacts with respect to typical mobility patterns, observed before the COVID-19 pandemic. 
The typical mean social contact value is computed by taking the average of daily peak social contact values observed on weekdays in January 2020. 
Thus, the social contact index ($SCI$) of 1 corresponds to the same amount of social contacts as the daily peak times on weekdays before the COVID-19 pandemic.

\bibliography{sample}

\end{document}